\documentclass[prl,nofootinbib,twocolumn,superscriptaddress]{revtex4}
\def\mysection#1{{\bf #1.} }
\def\mysections#1{{\bf #1.} }
\usepackage{amssymb}
\usepackage{amsmath}
\usepackage[dvips]{graphicx}
\usepackage{longtable}
\usepackage{verbatim}
\usepackage{amsfonts}

\newcommand{\be}{\begin{equation}}
\newcommand{\ee}{\end{equation}}
\newcommand{\bea}{\begin{eqnarray}}
\newcommand{\eea}{\end{eqnarray}}
\newcommand{\beq}{\begin{equation}}
\newcommand{\eeq}{\end{equation}}
\def\beqa{\begin{eqnarray}}
\def\eeqa{\end{eqnarray}}
\newcommand{\no}{\nonumber}
\def\lsim{\mathrel{\rlap{\lower4pt\hbox{\hskip1pt$\sim$}}
    \raise1pt\hbox{$<$}}}         
\def\gsim{\mathrel{\rlap{\lower4pt\hbox{\hskip1pt$\sim$}}
    \raise1pt\hbox{$>$}}}         

\begin{document}


\vspace*{-30mm}

\title{\boldmath Testing New Indirect CP Violation}

\author{Yuval Grossman}\email{yg73@cornell.edu}
\affiliation{Institute for High Energy Phenomenology, Newman
  Laboratory of Elementary Particle Physics, Cornell University,
  Ithaca, NY 14853, USA}

\author{Yosef Nir}\email{yosef.nir@weizmann.ac.il}
\affiliation{Department of Particle Physics,
  Weizmann Institute of Science, Rehovot 76100, Israel}

\author{Gilad Perez}\email{gilad.perez@weizmann.ac.il}
\affiliation{Department of Particle Physics,
  Weizmann Institute of Science, Rehovot 76100, Israel}

\vspace*{1cm}

\begin{abstract}
If new CP violating physics contributes to neutral meson mixing, but
its contribution to CP violation in decay amplitudes is negligible,
then there is a model independent relation between four (generally
independent) observables related to the mixing: The mass splitting
($x$), the width splitting ($y$), the CP violation in mixing
($1-|q/p|$), and the CP violation in the interference of decays with
and without mixing ($\phi$). For the four neutral meson systems,
this relation can be written in a simple approximate form:
$y\tan\phi\approx x(1-|q/p|)$. This relation is already tested
(successfully) in the neutral $K$ system. It provides predictions for
the $B_s$ and $D$ systems. The success or failure of these
relations will probe the physics that is responsible for the CP
violation.
\end{abstract}

\maketitle

\mysection{Introduction}
The fact that the Standard Model depends on a single CP violating phase
gives it a strong predictive power concerning CP asymmetries. The fact
that CP is a good symmetry of the strong interactions makes the
theoretical analysis of CP asymmetries often impressively clean. These
theoretical advantages, combined with the huge experimental progress
in the measurements of CP violation in $B$ decays and in the search
for CP violation in $B_s$ and $D$ decays, provide a powerful probe of
new physics. Observing deviations from the Standard Model predictions
will not only imply the existence of new physics, but also give
detailed information about features of the required new physics.

CP violation in meson decays can be classified to indirect and direct
CP violation. Indirect CP violation can be completely described by
phases in the dispersive part of the neutral meson mixing amplitude
($M_{12}$). In contrast, direct CP violation requires that there are
some phases in the decay amplitudes ($A_f$). Within the Standard
Model, many CP asymmetries require -- to an excellent approximation --
only indirect CP violation. Examples include $K\to\pi\pi$, $B\to\psi
K_S$ and $B_s\to\psi\phi$. This situation persists in many -- though
not all -- extensions of the Standard Model.

Indirect CP violation can manifest itself in two ways: CP violation
in mixing, which is the source of CP asymmetries in semileptonic
decays, and CP violation in the interference of decays with and
without mixing, which is often the dominant effect in decays into
final CP eigenstates. When there is no direct CP violation, these two
manifestations are not independent of each other. They are correlated
in a way that depends on the mass- and width-splittings between the
two neutral meson mass eigenstates. In this work, we derive this model
independent relation, and analyze its applicability and implications
in each of the four neutral meson systems ($K,D,B,B_s$).

\mysection{The experimental parameters}
We refer here explicitly to the neural $D$ system, but our
formalism applies equally well to all four neutral meson systems.
The two neutral $D$-meson mass eigenstates, $|D_1\rangle$ of mass $m_1$
and width $\Gamma_1$ and $|D_2\rangle$ of mass $m_2$ and width $\Gamma_2$,
are linear combinations of the interaction eigenstates $|D^0\rangle$ (with
quark content $c\bar u$) and $|\overline{D^0}\rangle$ (with quark content
$\bar cu$):
\beqa
|D_{1,2}\rangle&=&p|D^0\rangle\pm q|\overline{D^0}\rangle.
\eeqa
The average and the difference in mass and width are given by
\beqa
m\equiv\frac{m_1+m_2}{2},&\ \ \ \ &\Gamma\equiv\frac{\Gamma_1+\Gamma_2}{2},\no\\
x\equiv\frac{m_2-m_1}{\Gamma},&\ \ \ \ &y\equiv\frac{\Gamma_2-\Gamma_1}{2\Gamma}.
\eeqa
The decay amplitudes into a final state $f$ are defined as
$A_f=\langle f|{\cal H}|D^0\rangle$ and $\overline{A}_f=\langle
f|{\cal H}|\overline{D^0}\rangle$.
We define a complex dimensionless parameter $\lambda_f$:
\beq
\lambda_f=(q/p)(\overline{A}_f/A_f).
\eeq

As concrete examples, consider the doubly-Cabibbo-suppressed decay
$D^0\to K^+\pi^-$, the singly-Cabibbo-suppressed decay $D^0\to K^+K^-$,
and the Cabibbo-favored decay $D^0\to K^-\pi^+$.
Let us assume that effects of direct CP violation
are negligibly small even in the presence of new physics.
On the other hand, new physics could easily generate indirect
CP violation. The effects of indirect CP violation can be parameterized
in the following way:
\beqa\label{defphi}
\lambda^{-1}_{K^+\pi^-}&=& r_d|p/q|e^{-i(\delta_{K\pi}+\phi)},\no\\
\lambda_{K^-\pi^+}&=& r_d|q/p|e^{-i(\delta_{K\pi}-\phi)},\no\\
\lambda_{K^+K^-}&=&-|q/p|e^{i\phi},
\eeqa
where $r_d=|\overline{A}_{K^-\pi^+}/A_{K^-\pi^+}|$, $\delta_{K\pi}$
is a strong (CP conserving) phase, and $\phi$ is a weak
(CP violating) universal phase. The appearance of a single weak phase
that is common to all final states is related to the absence of direct
CP violation, while the absence of a strong phase in $\lambda_{K^+K^-}$
is related to the fact that the final state is a CP eigenstate.

We then have (see, for example \cite{Nir:2005js}), for $\Gamma t<1$,
\beqa\label{dcsdecay}
\frac{\Gamma[D^0(t)\to K^+\pi^-]}
{\Gamma[\overline{D^0}(t)\to K^+\pi^-]}
&=&r_d^2+r_d\left|\frac qp\right|
(y^\prime\cos\phi-x^\prime\sin\phi)\Gamma t\no\\
&&+\left|\frac qp\right|^2
\frac{y^2+x^2}{4}(\Gamma t)^2,
\no\\
\frac{\Gamma[\overline{D^0}(t)\to K^-\pi^+]}
{\Gamma[{D^0}(t)\to K^-\pi^+]}
&=&r_d^2+r_d\left|\frac pq\right|
(y^\prime\cos\phi+x^\prime\sin\phi)\Gamma t\no\\
&&+\left|\frac pq\right|^2\frac{y^2+x^2}{4}(\Gamma t)^2,
\eeqa
where $r_d=|\overline{A}_{K^-\pi^+}/A_{K^-\pi^+}|$,
$y^\prime=y\cos\delta_{K\pi}-x\sin\delta_{K\pi}$ and
$x^\prime=x\cos\delta_{K\pi}+y\sin\delta_{K\pi}$, and
\beqa\label{scsdecay}
\Gamma[D^0(t)&\to& K^+K^-]=e^{-\Gamma t}|A_{K^+K^-}|^2\no\\
&\times&[1-|q/p|(y\cos\phi-x\sin\phi)\Gamma t],\no\\
\Gamma[\overline{D^0}(t)&\to& K^+K^-]=e^{-\Gamma t}
|\overline{A}_{K^+K^-}|^2\no\\
&\times&[1-|p/q|](y\cos\phi+x\sin\phi)\Gamma t].
\eeqa

We focus our attention in this work on four parameters that are
related to $D^0-\overline{D^0}$ mixing: the two CP conserving
parameters $x$ and $y$ and the two CP violating parameters $(1-|q/p|)$
and $\phi$. It is clear from Eqs. (\ref{dcsdecay}) and
(\ref{scsdecay}) that, by fitting to the experimentally measured
time-dependent decay rates, one can extract these four parameters. We
thus call them ``experimental parameters''.

\mysection{The theoretical parameters}
The $\overline{D^0}-D^0$ transition amplitudes are defined as follows:
\beqa
\langle D^0|{\cal H}|\overline{D^0}\rangle&=&M_{12}-\frac i2\Gamma_{12},\no\\
\langle\overline{D^0}|{\cal H}|D^0\rangle&=&M_{12}^*-\frac i2\Gamma_{12}^*.
\eeqa
The overall phase of the mixing amplitude is not a physical
quantity. It can be changed by the choice of phase convention for
the up and charm quarks. The relative phase between $M_{12}$ and
$\Gamma_{12}$ is, however, phase convention independent and has
physics consequences.  The three physical quantities
related to the mixing can be defined as
\beq\label{thepar}
y_{12}\equiv|\Gamma_{12}|/\Gamma,\ \ \
x_{12}\equiv2|M_{12}|/\Gamma,\ \ \
\phi_{12}\equiv\arg(M_{12}/\Gamma_{12}).
\eeq
Given a particle physics model, one can calculate the three parameters
$y_{12},x_{12}$ and $\phi_{12}$ as a function of the model
parameters. We thus call them ``theoretical parameters''. Note that
$y_{12}$ is generated by final states that are common to $D^0$ and
$\overline{D^0}$ decays. Thus it is very likely that it is described
to a very good approximation by Standard Model physics (see, however,
\cite{Golowich:2006gq}). On the other hand, $x_{12}$ and $\phi_{12}$
can be affected by new physics parameters.

\mysection{From theory to experiment}
The following expressions give the experimental parameters in terms of
the theoretical ones:
\beqa\label{thexi}
x y&=&x_{12} y_{12}\cos\phi_{12},\no\\
x^2-y^2&=&x_{12}^2-y_{12}^2,\no\\
(x^2+y^2)\left|q/p\right|^2&=&
  x_{12}^2+y_{12}^2+2x_{12}y_{12}\sin\phi_{12},\no\\
x^2\cos^2\phi-y^2\sin^2\phi&=&x_{12}^2\cos^2\phi_{12}.
\eeqa
To obtain the last relation, we took into account the fact that, in
the absence of direct CP violation, we have for final CP eigenstates
\beq\label{nodirect}
{\cal I}m(\Gamma_{12}^* \overline{A}_f/A_f)=0,\ \ \
|\overline{A}_f/A_f|=1.
\eeq
The relations that we derive below depend crucially on this
condition. Even if, in general, there is direct CP violation in some
decays, our relations apply for those modes where
Eq. (\ref{nodirect}) holds.

We emphasize that the relation between the `theoretical' phase
$\phi_{12}$ (defined in Eq. (\ref{thepar})) and the `experimental'
phase $\phi$ (defined in Eq. (\ref{defphi})) is, in general, quite
complicated. In particular, when $x_{12}\lsim y_{12}$, as might still
be the case for the neutral $D$ system, the phase $\phi$ might be
considerably smaller than $\phi_{12}$ \cite{Bergmann:2000id}. In other
words, the new physics contribution could violate CP with a phase of
order one, yet $\phi$ is small.

\mysection{From experiment to theory}
Given experimental constraints on $x,y,|q/p|$ and $\phi$, we can use
Eq. (\ref{thexi}) to constrain $x_{12}$ and $\phi_{12}$ and
subsequently the new physics model parameters.  In particular, we
derived the following equations for each of $x_{12}$ and $\phi_{12}$,
first in terms of $x,y$ and $\phi$:
\beqa\label{xphiphi}
x_{12}^2&=&\frac{x^4\cos^2\phi+y^4\sin^2\phi}
{x^2\cos^2\phi-y^2\sin^2\phi},\\
\sin^2\phi_{12}&=&\frac{(x^2+y^2)^2\cos^2\phi\sin^2\phi}
{x^4\cos^2\phi+y^4\sin^2\phi},\no
\eeqa
and, second, in terms of $x,y$ and $|q/p|$:
\beqa\label{xphiqp}
x_{12}^2&=&x^2\frac{(1+|q/p|^2)^2}{4|q/p|^2}+
y^2\frac{(1-|q/p|^2)^2}{4|q/p|^2},\\
\sin^2\phi_{12}&=&\frac{(x^2+y^2)^2(1-|q/p|^4)^2}
{16x^2y^2|q/p|^4+(x^2+y^2)^2(1-|q/p|^4)^2}.\no
\eeqa

Let us assume, as is the case for $D$ decays at present, that $x$ and
$y$ are measured, while the CP violating parameters $(1-|q/p|)$ and
$\sin\phi$ are constrained to be small. For small $\sin\phi$ we
obtain, to ${\cal O}(\sin^2\phi)$,
\beqa\label{smallphi}
x_{12}^2&=&x^2\left[1+\frac{y^2(y^2+x^2)}{x^4}\sin^2\phi\right],\no\\
\sin^2\phi_{12}&=&\frac{(x^2+y^2)^2}{x^4}\sin^2\phi.
\eeqa
For small $(1-|q/p|)$ we obtain, to leading order:
\beqa\label{smallqp}
x_{12}^2&=&x^2\left[1+\frac{x^2+y^2}{x^2}
  \left(1-\left|\frac{q}{p}\right|\right)^2\right],\no\\
\sin^2\phi_{12}&=&\frac{(x^2+y^2)^2}{x^2y^2}
\left(1-\left|\frac qp\right|\right)^2.
\eeqa

\mysection{A model independent relation}
The fact that we are able to express the four experimental parameters
in terms of three theoretical ones means that the experimental
parameters fulfill a model independent relation. It depends solely on
our assumption that direct CP violation can be neglected.

The relation can be extracted from Eqs. (\ref{xphiphi}) and
(\ref{xphiqp}):
\beqa\label{moinre}
&&\frac{(1-|q/p|^4)^2}{\sin^2\phi}=\\
&&\ \frac{16(y/x)^2|q/p|^4+[1+(y/x)^2]^2(1-|q/p|^4)^2}
{1+(y/x)^4\tan^2\phi}\no.
\eeqa

The relation becomes very simple in two limits. Fortunately, each of
the four neutral meson systems is subject to at least one of these two
approximations. First, consider a system where
\beq
y_{12}\ll x_{12}.
\eeq
This approximation applies to the $B$ and $B_s$ systems.
It gives, to leading order in $y_{12}/x_{12}$:
\beqa\label{smallyot}
y/x&=&\cos\phi_{12}\ y_{12}/x_{12},\\
\left|q/p\right|-1&=&(y_{12}/x_{12})\sin\phi_{12},\ \
\tan\phi=-\tan\phi_{12}.\no
\eeqa
The derivation of the
sign for the CP violating observables starts from the definition of
$q/p$ (see, for example, \cite{akagan}).

Second, consider a system where CP violation is small,
\beq
|\sin\phi_{12}|\ll1.
\eeq
This situation applies to the $K$ system. Very recent measurements
imply that it also applies (with limits of order 0.2) to the $D$
system \cite{Barberio:2008fa}. We obtain, to leading order in
$|\sin\phi_{12}|$,
\beqa\label{smallphi}
y/x&=&{\rm sign}(\cos\phi_{12})\ y_{12}/x_{12},\\
\left|q/p\right|-1&=&\frac{(y/x)\tan\phi_{12}}{1+(y/x)^2},\ \
\tan\phi=\frac{-\tan\phi_{12}}{1+(y/x)^2}.\no
\eeqa

The two sets of equations, (\ref{smallyot}) and (\ref{smallphi}),
lead to the same simple relation:
\beq\label{phiqpbd}
\frac{y}{x}=\frac{1-|q/p|}{\tan\phi}.
\eeq
Eq. (\ref{phiqpbd}) is the main theoretical
result of this work. If it is found to be violated, then new
physics will have to provide not only indirect CP violation, but also
direct one. That would exclude many classes of candidate theories.

In what follows, we analyze the applicability and implications of
this relation in each of the four neutral meson systems.

\mysection{$K^0-\overline{K^0}$ mixing}
The two ingredients that go into the relation (\ref{phiqpbd}) -- small
CP violation and the absence of direct CP violation -- hold in the
$K\to\pi\pi$ decays. Thus, this relation should hold in the neutral
$K$ system. Neglecting direct CP violation, and defining
\beq
A_0=\langle (\pi\pi)_{I=0}|{\cal H}|K^0\rangle,\ \ \
\lambda_0=(q/p) (\bar A_0/A_0),
\eeq
the CP violating $\epsilon$ parameter corresponds to
\cite{Nir:1992uv}
\beq
\epsilon=\frac{1-\lambda_0}{1+\lambda_0}.
\eeq
Then we have
\beq
{\cal R}e(\epsilon)\approx\frac12(1-|q/p|),\ \ \
{\cal I}m(\epsilon)\approx-\frac12\tan\phi.
\eeq
The relation (\ref{phiqpbd}) translates into the prediction
\beq
\arg(\epsilon)\approx\arctan(-{x}/{y})=43.5^o,
\eeq
where, for the numerical value, we used \cite{Amsler:2008zzb}
$\Delta m_K=0.5290\times10^{10}\ {\rm s}^{-1}$ and
$\Delta\Gamma_K=-1.1163\times10^{10}\ {\rm s}^{-1}$. Indeed, the
experimental value is \cite{Amsler:2008zzb}
\beq
\arg(\epsilon)=43.51\pm0.05^o.
\eeq
Thus, the relation (\ref{phiqpbd}) is tested in the neutral kaon
system and works very well.

\mysection{$B^0-\overline{B^0}$ mixing}
In the neutral $B$ system, the width difference is constrained to be
small (and consistent with zero within the present accuracy),
$\Delta\Gamma/\Gamma=0.01\pm0.04$,  while the mass splitting is
measured to be much larger, $\Delta m/\Gamma=0.78\pm0.01$
\cite{Amsler:2008zzb}. Thus $y_{12}/x_{12}\ll1$ and Eqs. (\ref{smallyot})
apply. One has to note, however, that the equation for $\phi$ holds only
for modes where Eq. (\ref{nodirect}) applies. Since
\cite{Beneke:1996gn,Dighe:2001sr,Laplace:2002ik,Beneke:2003az,Lenz:2006hd}
\beq\label{gotbd}
\arg(\Gamma_{12})\approx\arg[(V_{tb}V_{td}^*)^2],
\eeq
the phase $\phi$ relates to modes whose phase is dominated
by $\arg(V_{tb}V_{td}^*)$. (The weak phase of $B\to\psi K_S$
is dominated by $\arg(V_{cb}V_{cd}^*)$ and, therefore, $S_{\psi K_S}$
cannot be used to test (\ref{phiqpbd}).) The problem is that the
approximation (\ref{gotbd}) gives $1-|q/p|=0$ and $\phi=0$, so that
$y\tan\phi=x(1-|q/p|)$ is fulfilled in a rather trivial way.

If one wants to go beyond (\ref{gotbd}), the large relative phase
between $V_{tb}V_{td}^*$ and $V_{cb}V_{cd}^*$ has to be taken into
account. It enters $\Gamma_{12}$ and $\overline{A}_f/A_f$ in different
ways, and thus direct CP violation plays a role and (\ref{phiqpbd}) is
violated. Nevertheless, the relation (\ref{phiqpbd}) could in
principle provide interesting predictions if $M_{12}$ had significant
contributions from new physics carrying a new phase. Experimental data
constrain, however, such contributions to be smaller than ${\cal
  O}(0.2)$ \cite{Bona:2007vi,ckmfitter}, which is the same order as
the direct CP violating effects in $\Gamma_{12}$
\cite{Beneke:1996gn,Dighe:2001sr,Laplace:2002ik,Beneke:2003az,Lenz:2006hd}.

\mysection{$B_s-\overline{B_s}$ mixing}
Within the Standard Model, the discussion the $B_s$ system follows
a line of reasoning that is very similar to our discussion of the
$B_d$ system. However, in contrast to the $B_d$ system, a situation
where the indirect CP violation is entirely dominated by new physics
in $M_{12}$ is still possible for $B_s-\overline{B_s}$ mixing. Actually,
recent measurements in D0 and CDF provide hints at a level higher
than $2\sigma$ that this is indeed the case \cite{Barberio:2008fa}. If
so, then the relation (\ref{phiqpbd}) provides a very interesting
probe of the new physics. Neglecting
$\beta_s=\arg[-(V_{ts}V_{tb}^*)/(V_{cs}V_{cb}^*]$, the relation reads
\beqa\label{aslspp}
A_{\rm SL}^s&=&-{\rm
  sign}(\cos\phi)(2y/x)S_{\psi\phi}/(1-S_{\psi\phi}^2)^{1/2}\no\\
&=&-2|y/x|\ S_{\psi\phi}/(1-S_{\psi\phi}^2)^{1/2}
\eeqa
where $A_{\rm SL}^s$ is the CP asymmetry in semileptonic decays,
and $S_{\psi\phi}$ is the CP violating parameter in the decays
into $(\psi\phi)_{CP=+}$. The second equality assumes
that neither $\Gamma_{12}$ nor $b\to c\bar cs$ decays are
significantly affected by new  physics, which implies that ${\rm
  sign}(y\cos\phi)={\rm sign}(y\cos\phi)^{\rm SM}=+1$.  
The experimental data read \cite{Amsler:2008zzb}
$\Delta\Gamma/\Gamma=-0.07\pm0.06,\ \Delta m/\Gamma=26.1\pm0.5$,
which give
\beq
y/x=-0.0014\pm0.0012.
\eeq
If the central value is approximately correct, then
$S_{\psi\phi}={\cal O}(0.3)$ would imply $A_{\rm SL}^s={\cal
  O}(-10^{-3})$. We can expect a significant improvement
in the measurements of $y$ and of $S_{\psi\phi}$. (Hopefully, the
hints for a signal in $S_{\psi\phi}$ will not disappear as the
experimental accuracy improves.) Then, we will obtain a much
sharper prediction for $A_{\rm SL}^s$. A failure of this test would
imply that the new physics introduces both direct and indirect
CP violation.

A relation very similar to (\ref{aslspp}) was previously presented in
Refs. \cite{Ligeti:2006pm,Blanke:2006ig}. Their relation can be written
as $A_{\rm SL}^s/S_{\psi\phi}={\cal R}e(\Gamma_{12}^{\rm SM}/M_{12}^{\rm
  SM})|M_{12}^{\rm SM}/M_{12}|$.
What we add here to their results are the following two points:
\begin{enumerate}
\item The right hand side of this relation,
which is calculated from theory, can be replaced by the experimentally
measurable factor $-2y/(x\cos\phi)$. Thus, this becomes a theory-independent (in both
the electroweak model and QCD uncertainty aspects) relation.
\item We make it clear that a failure of this relation must imply
new direct CP violation.
\end{enumerate}

\mysection{$D^0-\overline{D^0}$ mixing}
Within the Standard Model, CP violation in $D^0-\overline{D^0}$
is negligibly small (see, for example, \cite{Blaylock:1995ay}).
Thus, any signal of CP violation requires new physics. It is
quite likely that such new physics will contribute negligibly
to tree level decay amplitudes, though new direct CP violation
is not impossible \cite{Grossman:2006jg}. Measurements of the
time dependent decay rates (\ref{dcsdecay}) and (\ref{scsdecay})
will allow us to extract $\phi$ and $1-|q/p|$ and put (\ref{phiqpbd})
to the test.

Experimentally, there has been a very significant progress in
determining the mixing parameters in the neutral $D$ system
\cite{Barberio:2008fa}:
\beqa
x&=&(1.00\pm0.25)\times10^{-2},\no\\
y&=&(0.77\pm0.18)\times10^{-2},\no\\
1-|q/p|&=&+0.06\pm0.14,\no\\
\phi&=&-0.05\pm0.09.
\eeqa
The CP violating parameters are constrained to be small, and
consistent with zero. In case, however, that CP violation is
observed in the future, the fact that
\beq
y/x\approx0.8\pm0.3
\eeq
suggests that the CP violation in mixing is comparable in size to
the CP violation in the interference of decays with and without
mixing. Whether or not the relation (\ref{phiqpbd}) is fulfilled
will teach us about the new physics and will disfavor or support
models of the type discussed in Ref.  \cite{Grossman:2006jg},
where direct CP violation can be generated.

\mysection{Conclusions}
CP asymmetries in neutral meson decays where direct CP violation
is negligible obey a relation. The relation involves four
experimentally measurable parameters and is thus independent of the
electroweak model and clean of QCD uncertainties. It applies to neutral
$K$ and $D$ decays in the form (\ref{phiqpbd}). If new physics provides
a large phase to $B_s-\overline{B_s}$ mixing, then the same relation
applies also to $B_s$ decays.

The phenomenological implications of this relation are the following:
\begin{itemize}
\item The relation is already successfully tested in $K$ decays.
\item If a large CP violating effect is measured in $B_s\to\psi\phi$,
then there is a clear prediction for the CP asymmetry in semileptonic
decays $A_{\rm SL}^s$ that is strongly enhanced compared to the SM.
\item If, for neutral $D$ decays, CP violation in either mixing or
the interference of decays with and without mixing is observed,
there is a clear prediction for CP violation of the other type,
of comparable size.
\item If the relation fails in $D$ decays, it will be an unambiguous
evidence that the new physics generates also CP violation in the
decay amplitudes.
\end{itemize}

\mysections{Acknowledgments}
We are grateful to Alex Kagan for pointing out sign errors in the
first version of this paper. We thank Monika Blanke for pointing out a 
missing factor of 2 in Eq. (\ref{aslspp}) \cite{Bigi:2009df}. This work is
supported by the United States-Israel Binational Science Foundation
(BSF), Jerusalem, Israel.  The work of YG is supported by the NSF
grant PHY-0757868. The work of YN is supported by the Israel Science
Foundation founded by the Israel Academy of Sciences and Humanities, 
the German-Israeli foundation for scientific research and development
(GIF), and the Minerva Foundation. The work of GP is supported by the
Peter and Patricia Gruber Award.


\end{document}